\documentclass[fleqn,usenatbib]{mnras}

\usepackage{newtxtext,newtxmath}

\usepackage[T1]{fontenc}

\usepackage{CJK}
\usepackage{color}
\usepackage[svgnames]{xcolor}
\usepackage{soul}
\usepackage{amsmath} 
\usepackage{bm} 
\usepackage{mathrsfs}
\usepackage{wasysym}
\usepackage{amsmath}
\usepackage{threeparttable}
\usepackage{multirow} 
\usepackage{array}
\usepackage{lineno} 
\usepackage[normalem]{ulem} 

\DeclareRobustCommand{\VAN}[3]{#2}
\let\VANthebibliography\thebibliography
\def\thebibliography{\DeclareRobustCommand{\VAN}[3]{##3}\VANthebibliography}

\usepackage{graphicx}	
\usepackage{amsmath}	

\title[Constraining H$_0$ with Neural Networks]{Constraining the Hubble Constant with a Simulated Full Covariance Matrix Using Neural Networks}

\author[Jing Niu et al.]{
Jing Niu (牛菁),$^{1,2}$
Peng He (贺鹏),$^{3}$
Tong-Jie Zhang (张同杰),$^{1,2}$\thanks{E-mail: tjzhang@bnu.edu.cn}
\\
$^{1}$Institute for Frontiers in Astronomy and Astrophysics, Beijing Normal University, Beijing 102206, China\\
$^{2}$School of Physics and Astronomy, Beijing Normal University, Beijing 100875, China\\
$^{3}$Burerau of Frontier Science and Education, Chinese Academy of Sciences, Beijing 100190, China
}

\begin{document}


\begin{CJK*}{UTF8}{gbsn}
\label{firstpage}
\pagerange{\pageref{firstpage}--\pageref{lastpage}}
\maketitle

\begin{abstract}
The Hubble parameter, $H(z)$, plays a crucial role in understanding the expansion history of the universe and constraining the Hubble constant, $\mathrm{H}_0$. The Cosmic Chronometers (CC) method provides an independent approach to measuring $H(z)$, but existing studies either neglect off-diagonal elements in the covariance matrix or use an incomplete covariance matrix, limiting the accuracy of $\mathrm{H}_0$ constraints. To address this, we use a Positive-Definite Covariance Network (PD-CovNet) to simulate the full $33 \times 33$ covariance matrix based on a previously published $15 \times 15$ covariance matrix. Hyperparameters are chosen via leave-one-z-out validation, and performance is benchmarked against a Gaussian-process (GP) baseline. Under identical five-fold cross-validation over redshift groups, we prove that PD-CovNet is a reliable generator of the full covariance compared to the GP baseline. Using this full PD-CovNet-simulated covariance alongside three comparators with different covariance specifications, we constrain \(\mathrm{H}_0\) with two independent methods (EMCEE and GP). Across all covariance specifications and both constraint methods, standardized differences and two-sided p-values show no statistically meaningful shift in the central value of the constrained \(\mathrm{H}_0\). However, the precision of the constrained \(\mathrm{H}_0\) depends on both covariance and method: EMCEE is uniformly more precise than GP once covariance is modeled; within a fixed method, incorporating more covariance reduces precision; and PD-CovNet hyperparameters have a modest effect on uncertainty. These results indicate the importance of accurate covariance modeling in CC-based \(\mathrm{H}_0\) constraints.

\end{abstract}




\section{Introduction}
\label{sec:intro}
The study of the universe's evolution is a fundamental and highly significant astronomy research area. The Hubble parameter, $H(z)$, describes the expansion rate of the universe at a given redshift $z$, where a higher $z$ corresponds to an earlier epoch in the universe's history. Accurate measurements of $H(z)$ are crucial for understanding the dynamics of cosmic expansion and the nature of dark energy. Various methods have been developed to measure $H(z)$, including Type Ia Supernovae (SNe Ia) \citep{1998AJ....116.1009R, 1999ApJ...517..565P}, Baryon Acoustic Oscillations \citep{2005ApJ...633..560E, 2017MNRAS.470.2617A}, and Cosmic Chronometers (CC) \citep{2002ApJ...573...37J}, among others. The $H(z)$ dataset can be used to constrain the Hubble constant, $\mathrm{H}_0$, which represents the present-day expansion rate of the universe. Two key approaches to measuring $\mathrm{H}_0$, the analysis of cosmic microwave background (CMB) anisotropies \citep{2020A&A...641A...6P} and the distance ladder method using Cepheids and SNe Ia \citep{2022ApJ...934L...7R}, yield significantly different results. The discrepancy between these measurements exceeds $5\sigma$ \citep{2019NatAs...3..891V, 2020NatRP...2...10R}, a conflict commonly referred to as the Hubble tension. The CC method offers an independent, third approach to investigating this issue.

To achieve a more accurate determination of $\mathrm{H}_0$ using the CC dataset, two key approaches can be considered: obtaining additional $H(z)$ data points and incorporating the covariance matrix between the data points. A pioneering study by \citet{2020ApJ...898...82M} calculated a $15 \times 15$ covariance matrix for the $H(z)$ data points they previously measured. However, a full covariance matrix for the entire CC dataset has not been proposed. Different studies address this issue in varying ways: (1) the most common approach assumes no covariance between the data points \citep{2023PDU....3901147N, 2023ApJS..266...27Z}, and (2) some studies use only the published $15 \times 15$ covariance matrix and assume that all other off-diagonal elements in the full covariance matrix are zero \citep{2024ApJS..270...23Z, 2024ApJ...972...14N}. These approaches to handling the covariance matrix are insufficient to fully evaluate how the full covariance matrix affects the constrained $\mathrm{H}_0$ value or to quantify the extent of its impact.

To address this problem, we use the Positive-Definite Covariance Network (PD-CovNet) to expand the published \(15\times15\) covariance of \citet{2020ApJ...898...82M} to a full \(33\times33\) covariance for the CC dataset. Because PD-CovNet-simulated covariances are sensitive to hyperparameter choice, we select the hyperparameters via leave-one-z-out (LOO) cross-validation. As a nonparametric baseline, we also train a GP model to generate the \(33\times33\) covariance and compare it to PD-CovNet via five-fold cross-validation. To minimize the influence of different methods, we adopt two independent approaches, the Affine Invariant Markov Chain Monte Carlo Ensemble Sampler (EMCEE) \citep{2013PASP..125..306F} and Gaussian Process (GP) regression \citep{2006gpml.book.....R, 2012JCAP...06..036S}, to constrain $\mathrm{H}_0$. With PD-CovNet tuned via LOO, we simulate the full covariance and quantify the impact of both the covariance specification and the constraint method on the $\mathrm{H}_0$ estimate and its uncertainty.

This article is organized as follows. Section~\ref{sec:Data} describes the CC data and reports the published \(15\times15\) covariance. Section~\ref{sec:Simulate Covariance Matrix} introduces PD-CovNet and simulates a full \(33\times33\) covariance. It summarizes the architecture, selects hyperparameters, and compares the PD-CovNet-simulated covariance with a Gaussian-process baseline. Section~\ref{sec:Constraining H0 Using the CC Dataset} then constrain \(\mathrm{H}_0\) with two independent methods, EMCEE and GP, and assesses how the covariance specification and constraint method affect the $\mathrm{H}_0$ estimate and its uncertainty. Section~\ref{sec:Conclusion} concludes.

\section{Data} \label{sec:Data}
The CC approach offers a model-independent method for directly determining $H(z)$ values. This method, known as the differential age method, is grounded in the fundamental definition of the Hubble parameter:
\begin{equation}
H(z) = -\frac{1}{1+z} \frac{\mathrm{d}z}{\mathrm{d}t}.
\label{eq:data1}
\end{equation}
By measuring $\mathrm{d}z/\mathrm{d}t$ through observations of massive, passive galaxies, the CC method offers an independent way to explore the universe's expansion history without requiring assumptions about the cosmological model \citep{2002ApJ...573...37J, 2020ApJ...898...82M, 2023ApJS..265...48J}. The CC dataset used in this study is presented in Table~\ref{tab:cc_data}.

\begin{table}
    \centering
    \caption{Compiled CC data}
    \label{tab:cc_data}
    \begin{tabular}{ccc}
        \hline
        Redshift $z$ & $H(z)$ \textsuperscript{a} $\pm 1\sigma$ error & References \\
        \hline
        0.07 & $69 \pm 19.6$ & \cite{2014RAA....14.1221Z} \\
        0.09 & $70.7 \pm 12.3$\textsuperscript{b} & \cite{2003ApJ...593..622J} \\
        0.12 & $68.6 \pm 26.2$ & \cite{2014RAA....14.1221Z} \\
        0.17 & $83 \pm 8$ & \cite{2005PhRvD..71l3001S} \\
        0.1791 & $75 \pm 4$ & \cite{2012JCAP...08..006M} \\
        0.1993 & $75 \pm 5$ & \cite{2012JCAP...08..006M} \\
        0.2 & $72.9 \pm 29.6$ & \cite{2014RAA....14.1221Z} \\
        0.27 & $77 \pm 14$ & \cite{2005PhRvD..71l3001S} \\
        0.28 & $88.8 \pm 36.6$ & \cite{2014RAA....14.1221Z} \\
        0.3519 & $83 \pm 14$ & \cite{2012JCAP...08..006M} \\
        0.382 & $83 \pm 13.5$ & \cite{2016JCAP...05..014M} \\
        0.4 & $95 \pm 17$ & \cite{2005PhRvD..71l3001S} \\
        0.4004 & $77 \pm 10.2$ & \cite{2016JCAP...05..014M} \\
        0.4247 & $87.1 \pm 11.2$ & \cite{2016JCAP...05..014M} \\
        0.4497 & $92.8 \pm 12.9$ & \cite{2016JCAP...05..014M} \\
        0.47 & $89 \pm 49.6$ & \cite{2017MNRAS.467.3239R} \\
        0.4783 & $80.9 \pm 9$ & \cite{2016JCAP...05..014M} \\
        0.48 & $97 \pm 62$ & \cite{2010JCAP...02..008S} \\
        0.5929 & $104 \pm 13$ & \cite{2012JCAP...08..006M} \\
        0.6797 & $92 \pm 8$ & \cite{2012JCAP...08..006M} \\
        0.7812 & $105 \pm 12$ & \cite{2012JCAP...08..006M} \\
        0.8 & $113.1 \pm 25.22$ & \cite{2023ApJS..265...48J} \\
        0.8754 & $125 \pm 17$ & \cite{2012JCAP...08..006M} \\
        0.88 & $90 \pm 40$ & \cite{2010JCAP...02..008S} \\
        0.9 & $117 \pm 23$ & \cite{2005PhRvD..71l3001S} \\
        1.037 & $154 \pm 20$ & \cite{2012JCAP...08..006M} \\
        1.26 & $135 \pm 65$ & \cite{2023AA...679A..96T} \\
        1.3 & $168 \pm 17$ & \cite{2005PhRvD..71l3001S} \\
        1.363 & $160 \pm 33.6$ & \cite{2015MNRAS.450L..16M} \\
        1.43 & $177 \pm 18$ & \cite{2005PhRvD..71l3001S} \\
        1.53 & $140 \pm 14$ & \cite{2005PhRvD..71l3001S} \\
        1.75 & $202 \pm 40$ & \cite{2005PhRvD..71l3001S} \\
        1.965 & $186.5 \pm 50.4$ & \cite{2015MNRAS.450L..16M} \\
        \hline
    \end{tabular}
    \vspace{3mm}
    \parbox{\linewidth}{\small
    \textsuperscript{a} $H(z)$ figures are in the unit of km s$^{-1}$ Mpc$^{-1}$.\\
    \textsuperscript{b} We correct the value of $H(0.09)$ in Appendix \ref{appendix: The correction of H(0.09)}.
    }
\end{table}

Recent studies suggest that the systematic uncertainties in the observed H(z) values within the CC dataset have not been adequately addressed. \citet{2020ApJ...898...82M} conducted a detailed analysis of the components contributing to these uncertainties. Their research identifies four primary sources of systematic uncertainty: the stellar population synthesis (SPS) model, the metallicity of the stellar population, the star formation history (SFH), and residual star formation from a young subdominant stellar component. The covariance matrix associated with the CC method can be expressed as \citep{2020ApJ...898...82M}:
\begin{equation}
\mathrm{Cov}_{ij} = \mathrm{Cov}_{ij}^{\mathrm{stat}} + \mathrm{Cov}_{ij}^{\mathrm{young}} + \mathrm{Cov}_{ij}^{\mathrm{model}} + \mathrm{Cov}_{ij}^{\mathrm{met}}.
\label{eq:data2}
\end{equation}
The covariance matrix, $\mathrm{Cov}_{ij}$, between the Hubble parameter values at redshifts $z_i$ and $z_j$ incorporates errors from four distinct sources: statistical errors ("stat"), residual star formation from a young subdominant component ("young"), the choice of model ("model"), and stellar metallicity ("met"). Specifically, the "model" component can be further divided into subcategories, including SFH, initial mass function (IMF), stellar library (st.lib), and SPS model,
\begin{equation}
\mathrm{Cov}_{ij}^{\mathrm{model}} = \mathrm{Cov}_{ij}^{\mathrm{SFH}} + \mathrm{Cov}_{ij}^{\mathrm{IMF}} + \mathrm{Cov}_{ij}^{\mathrm{st.lib}} + \mathrm{Cov}_{ij}^{\mathrm{SPS}}.
\label{eq:data3}
\end{equation}
Among the contributions to the systematic error, $\mathrm{Cov}_{ij}^{\mathrm{met}}$ in Equation~(\ref{eq:data2}) and $\mathrm{Cov}_{ij}^{\mathrm{SFH}}$ have already been accounted for in the published CC data presented in Table~\ref{tab:cc_data} by \citet{2012JCAP...08..006M}, \citet{2016JCAP...05..014M}, and \citet{2015MNRAS.450L..16M}. Residual contamination from a young stellar population, denoted as $\mathrm{Cov}_{ij}^{\mathrm{young}}$, contributes negligibly to the systematic error \citep{2018ApJ...868...84M}.

Excluding the contributions analyzed above, the systematic error in Equations (\ref{eq:data2}) and (\ref{eq:data3}) consists of three remaining components: $\mathrm{Cov}_{ij}^{\mathrm{SPS}}$, $\mathrm{Cov}_{ij}^{\mathrm{st.lib}}$, and $\mathrm{Cov}_{ij}^{\mathrm{IMF}}$. \citet{2020ApJ...898...82M} analyzed these components in detail, considering 12 combination models involving various SPS models, stellar libraries, and IMFs, with each combination denoted as a specific model (e.g., $a$, $b$, etc.). The covariance matrix, $\mathrm{Cov}_{ij}^{\mathrm{model}}$, quantifies the systematic error of Hubble parameters between redshifts $z_i$ and $z_j$ caused by these components and is defined as:
\begin{equation}
\mathrm{Cov}_{ij} = \widehat{\eta}(z_i) \times H(z_i) \times \widehat{\eta}(z_j) \times H(z_j).
\label{eq:data4}
\end{equation}
Where $\widehat{\eta}(z_i)$ represents the mean percentage bias,
\begin{equation}
\eta(z)_{a,b} = \frac{H(z)_{a,b}-H_{fid}(z)}{H_{fid}(z)},
\label{eq:data5}
\end{equation}
\begin{equation}
\widehat{\eta}(z) = \mathrm{mean}(\mathrm{abs}(\eta(z)_{a,b})).
\label{eq:data6}
\end{equation}
Here, the Hubble parameter $H(z)_{a,b}$ is generated based on models $a$ and $b$, while $\eta(z)_{a,b}$ represents the percentage bias matrix for $H(z)$. To quantify all contributions to the systematic error matrix, the mean percentage bias, $\widehat{\eta}(z_i)$, is calculated as the absolute mean across 12 combinations of various SPS models, stellar libraries, and IMFs. The detailed methodology is described in \citet{2020ApJ...898...82M}, and the source code they used is publicly accessible.\footnote{\href{https://gitlab.com/mmoresco/CCcovariance/-/tree/master}{https://gitlab.com/mmoresco/CCcovariance/-/tree/master}}

\section{Simulate Covariance Matrix} \label{sec:Simulate Covariance Matrix}
\citet{2020ApJ...898...82M} calculated the $15 \times 15$ covariance matrix associated with the systematic error using Equation~(\ref{eq:data4}). However, the full covariance matrix for the 33 $H(z)$ CC data points listed in Table~\ref{tab:cc_data} remains unknown, making its impact on the constrained value of $\mathrm{H}_0$ using the CC dataset unclear. In Section \ref{subsec:PD-CovNet}, we simulate the full covariance matrix from the published $15\times15$ covariance matrix using the PD-CovNet. Section \ref{subsec:GP baseline} describes a GP baseline trained on the same published covariance matrix. We then compare the simulated $33 \times 33$ covariance matrices from both methods and report the results in Section \ref{subsec:Five-fold cross-validation}.

\subsection{PD-CovNet}
\label{subsec:PD-CovNet}
Our goal is to construct a positive-definite (PD) covariance matrix over the redshift positions by learning from the published $15\times15$ covariance matrix. In this section we introduce PD-CovNet, an artificial neural network (ANN)-based positive-definite covariance model. ANNs are inspired by biological neural systems and learn by adjusting connection weights during training \citep{1943Bulletin.....5M}. Unlike traditional methods that rely on predefined functional forms or handcrafted features, ANNs learn flexible nonlinear representations from data. We learn a feature map $\phi(z)$ via a Fully Connected Neural Network (FCNN), mapping each redshift to an $r$-dimensional representation (Section~\ref{subsubsec:Model Architecture}). The covariance is then built from these features and is positive definite by construction. In Section~\ref{subsubsec:selection of hyperparameters}, we select hyperparameters to obtain the model that best fits the published $15\times15$ covariance matrix, and then use it to simulate the full covariance matrix for the CC data in Table~\ref{tab:cc_data}.

\subsubsection{Model Architecture}
\label{subsubsec:Model Architecture}
Let $z_{1:m}$ be the training redshifts and let $S\in \mathbb{R}^{m\times m}$ denote the published covariance on this grid (in our study, $m=15$). To keep data and model distinct, we use $S$ for the published (empirical) matrix and $\Sigma$ for the model-simulated covariance; here $\mathbb{R}$ denotes real-valued vectors or matrices. Given any target grid $z_{1:N}$ (in our study, $N=33$), we learn parameters $\theta$ such that the model produces a symmetric, positive-definite $\Sigma_\theta(z_{1:N})\in \mathbb{R}^{N\times N}$. We standardize redshifts using the training statistics and apply the same standardization when evaluating any other redshift list.

We simulate the CC covariance matrix using PD-CovNet, which learns a feature map with an FCNN. FCNN is a fundamental architecture in deep learning, forming the basis for many advanced neural networks. FCNNs rely exclusively on densely connected layers where every neuron in one layer is connected to every neuron in the subsequent layer. This structure ensures flexibility and general applicability to various tasks. 

The primary objective of FCNN is to learn a mapping function $f: \mathbf{x} \to \hat{\mathbf{y}}$, where the input $\mathbf{x} = [x_1, x_2, \dots, x_m]$ is an $m$-dimensional feature vector, and the output $\hat{\mathbf{y}}$ is a vector of predicted values. This mapping is achieved through a sequence of layers, each consisting of linear transformations followed by non-linear activation functions. Mathematically, an FCNN can be expressed as a composite function:
\begin{equation}
    f(\mathbf{x}; \mathbf{\Theta}) = h_L(h_{L-1}(\dots h_2(h_1(\mathbf{x}))))),
    \label{eq:Model Architecture1}
\end{equation}
where $\mathbf{\Theta}$ denotes the parameters of the network, including the weights and biases across all layers. The transformation at layer $l$ is represented by $h_l$, and $L$ is the total number of layers in the network. For a specific layer $l$, the transformation is defined as:
\begin{equation}
    h_l(\mathbf{u}) = \sigma(\mathbf{W}_l \mathbf{u} + \mathbf{b}_l),  
    \label{eq:Model Architecture2}
\end{equation}
where $\mathbf{u}$ is the input to layer $l$, $\mathbf{W}_l$ is the weight matrix for layer $l$, $\mathbf{b}_l$ is the bias vector with dimension $n_l$, and $\sigma$ is an element-wise activation function that introduces non-linearity.

\begin{figure*}
    \centering
    \includegraphics[width=0.7\linewidth]{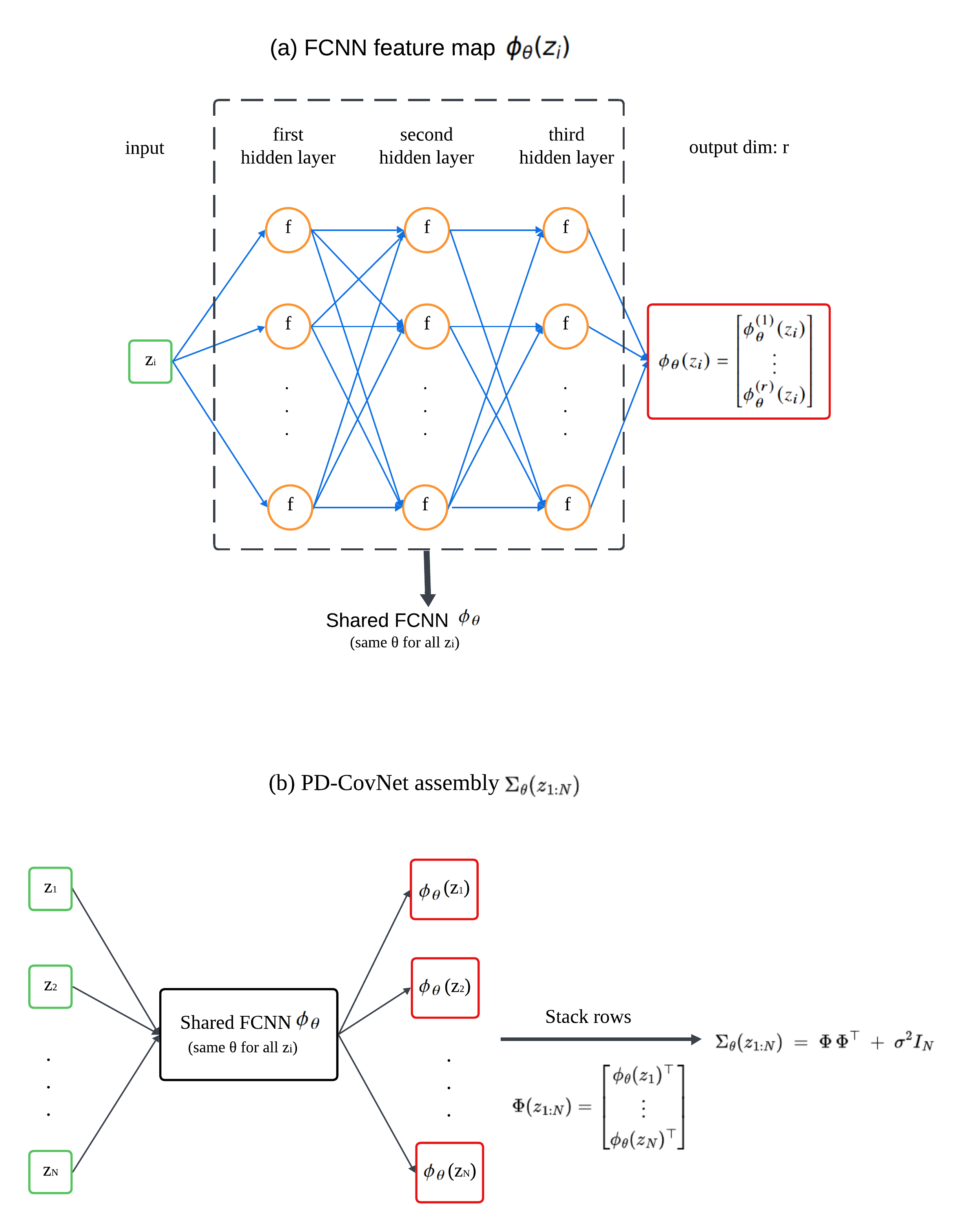} 
    \caption{Architecture of the shared FCNN feature map and PD-CovNet. (a) FCNN feature map $\phi_\theta(z_i)$: a fully connected network with three hidden layers (128 neurons each, employing the ReLU activation function) mapping each standardized $z_i\in\mathbb{R}$ to an $r$-dimensional feature vector $\phi_\theta(z_i)$; the same weights $\theta$ are used for all $i$. (b) PD-CovNet constructs a covariance over the redshift grid by applying the shared FCNN to $N$ inputs and stacking the outputs row-wise to form $\Phi_\theta(z_{1:N})$ in Equation~(\ref{eq:Model Architecture5}). The covariance $\Sigma_\theta(z_{1:N})$ is designed by Equation~(\ref{eq:Model Architecture6}). In this paper, parameters $(\theta,\sigma^2)$ are learned from the published $15\times15$ covariance via the multivariate normal log-likelihood and then used to simulate the full $33\times33$ covariance.}
    \label{fig:StructureFCNN_PDCovNet}
\end{figure*}

In this research, we use a shared FCNN to map each standardized redshift $z_i$ to an $r$-dimensional feature vector $\phi_\theta(z_i)$. The feature map is
\begin{equation}
    f: z_i \to \phi_\theta(z_i),
    \label{eq:Model Architecture3}
\end{equation}
where
\begin{equation}
    \phi_\theta(z_i)=\big(\phi_\theta^{(1)}(z_i),\ldots,\phi_\theta^{(r)}(z_i)\big)^\top.
    \label{eq:Model Architecture4}
\end{equation}
Panel (a) of Figure~\ref{fig:StructureFCNN_PDCovNet} illustrates the FCNN used in this study.


Given a target grid $z_{1:N}$, apply the same weight-shared network to each $z_i$. Stacking these row-wise forms the feature matrix
\begin{equation}
\Phi(z_{1:N}) =
\begin{bmatrix}
\phi_\theta(z_1)^\top\\
\vdots\\
\phi_\theta(z_N)^\top
\end{bmatrix}.
    \label{eq:Model Architecture5}
\end{equation}
We design the covariance over the $N$ redshifts as
\begin{equation}
\Sigma_\theta(z_{1:N}) \;=\; \Phi\,\Phi^\top \;+\; \sigma^2 I_N.
\label{eq:Model Architecture6}
\end{equation}
The covariance construction is illustrated in panel (b) of Figure~\ref{fig:StructureFCNN_PDCovNet}. Because $\Phi\Phi^\top$ is a Gram matrix, it is symmetric positive semidefinite. To ensure the multivariate normal log-likelihood is well defined and numerically stable, we add a diagonal term $\sigma^2 I_N$ with $\sigma^2>0$, which both models small-scale variance and guarantees strict positive definiteness. Here $I_N$ denotes the $N\times N$ identity matrix, where $N$ is the number of redshift points. $\sigma^2$ is a scalar noise variance learned from the data during training.

We estimate $\theta$ by maximizing the multivariate normal likelihood on the $15\times15$ published covariance matrix. Equivalently, we minimize the corresponding negative log-likelihood
\begin{equation}
\mathcal{L}(\theta)=\log\det\Sigma_\theta+\operatorname{tr}\!\big(\Sigma_\theta^{-1}S\big),
\label{eq:Model Architecture7}
\end{equation}
where $S$ is the empirical covariance computed from the $15\times15$ block. We evaluate $\mathcal{L}(\theta)$ using a stable factorization of $\Sigma_\theta$. Our parameterization guarantees that $\Sigma_\theta$ is strictly positive definite, so the factorization and the quantities $\log\det\Sigma_\theta$ and $\Sigma_\theta^{-1}$ are well defined and numerically stable. Gradients are obtained via automatic differentiation and optimized with Adam, which uses a base learning rate and automatically adjusts the step size for each parameter to keep training stable.

\begin{figure*}
    \centering
    \includegraphics[width=1\linewidth]{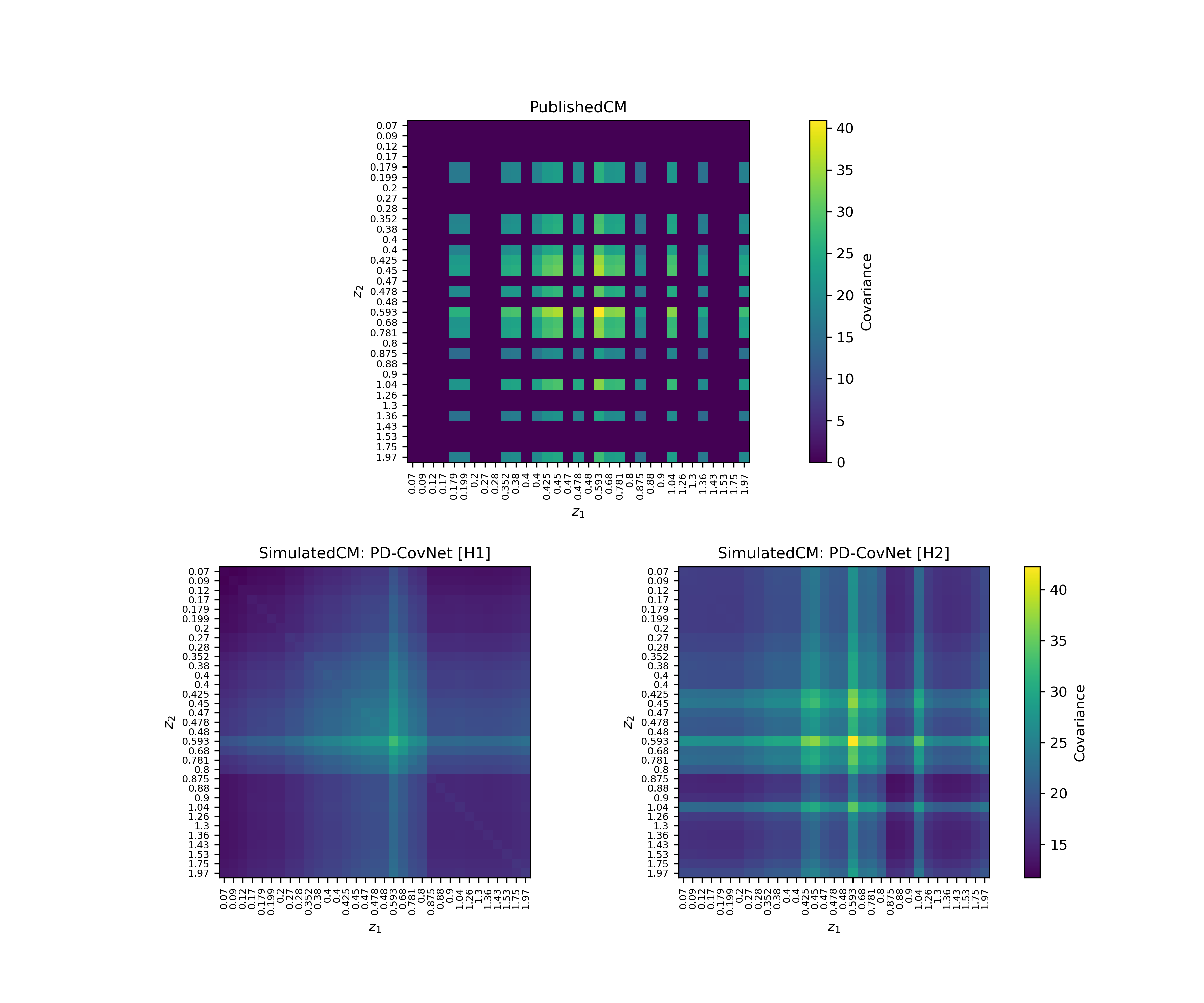} 
    \caption{The published covariance matrix (top) and PD-CovNet-simulated covariance matrices (bottom). The top panel shows the $15\times15$ covariance matrix reported by \protect\cite{2020ApJ...898...82M}; the remaining entries for the full set of 33 $H(z)$ CC data points (see Table~\ref{tab:cc_data}) are set to zero. The bottom panels are PD-CovNet simulations under two hyperparameter sets: H1 (bottom left), specified as epochs $=500$, feature dimension $=12$, hidden width $=32$, hidden depth $=2$, and learning rate $=3\times10^{-3}$; H2 (bottom right), specified as epochs $=3250$, feature dimension $=4$, hidden width $=128$, hidden depth $=3$, and learning rate $=10^{-3}$. The $x$- and $y$-axes are the redshifts $z_1$ and $z_2$, respectively. The top panel has its own colorbar; the two bottom panels share a common colorbar.}
    \label{fig:top_publishedCM_bottom_PDCovNetCMs}
\end{figure*}

We have now completed PD-CovNet for CC data. We train on the $15\times15$ published covariance block: using Equations (\ref{eq:Model Architecture3}), (\ref{eq:Model Architecture4}), and (\ref{eq:Model Architecture5}) to build the feature matrix on the 15-point grid, forming $\Sigma_{\theta}(z_{1:15})$ via Equation~(\ref{eq:Model Architecture6}), and minimizing the multivariate normal likelihood in Equation~(\ref{eq:Model Architecture7}) with Adam to learn $\theta$ and $\sigma^2$. With the trained parameters, we then simulate the full $33\times33$ covariance by evaluating the feature map on the 33-point target grid and returning $\Sigma_{\theta}(z_{1:33})$ as given by Equation~(\ref{eq:Model Architecture6}). We also simulate under different hyperparameter settings shown in Figure~\ref{fig:top_publishedCM_bottom_PDCovNetCMs}. Because the simulated covariance matrices vary noticeably with these choices, we adopt the hyperparameters selected in Section \ref{subsubsec:selection of hyperparameters} for the final simulation.

\subsubsection{selection of hyperparameters.}
\label{subsubsec:selection of hyperparameters}
Figure~\ref{fig:top_publishedCM_bottom_PDCovNetCMs} shows that hyperparameters materially affect the simulated \(33\times33\) covariance. The hyperparameters we selected in this paper and their option ranges are:
\begin{itemize}

\item \textbf{Epochs (milestones)} in range \([250,\,7000]\) with step \(250\). We checkpoint at each milestone and apply early stopping at the earliest milestone where the validation score no longer decreases, which prevents overfitting by stopping before validation performance deteriorates. Earlier epochs yield smoother, underfit simulated covariance matrices. Later epochs add detail but can introduce noise beyond the stopping point.

\item \textbf{Feature dimension \(r\)} in \(\{4,8,12\}\). It controls the effective rank of the learned feature covariance \(\Phi\Phi^\top\) through \(\phi_\theta\) in Equation~(\ref{eq:Model Architecture4}). Smaller \(r\) yields smoother simulated covariance matrices, whereas larger \(r\) preserves finer structure.

\item \textbf{Hidden width} (number of neurons per layer) in \(\{32,64,128\}\) and \textbf{Hidden depth} (number of hidden layers) in \(\{1,2,3\}\). These define the capacity of the shared feature network \(\phi_\theta(z)\) in Equation~(\ref{eq:Model Architecture3}). Wider or deeper models capture more detailed patterns and interactions in the published $15 \times 15$ training covariance block. However, if they are too large, they risk overfitting, producing spurious oscillations that are noise. Narrower or shallower models yield smoother simulated covariance matrices.

\item \textbf{Learning rate} in \(\{3\times10^{-4},\,10^{-3},\,3\times10^{-3}\}\). This is the Adam step size. Smaller values are more stable and may underfit, while larger values can speed training but may cause noisy updates and high-frequency artifacts.

\end{itemize}

We select PD-CovNet hyperparameters by minimizing the leave-one-z-out (LOO) median RMSE computed on the published $15 \times 15$ covariance matrix. LOO is a form of cross-validation in which, for a dataset of size $n$, we fit the same model $n$ times, each time leaving out exactly one unit for validation and training on the remaining $n-1$ units. Cycling through all units yields one validation score per left-out unit by comparing the model's prediction for that unit with its observed data. Aggregating these scores estimates generalization while making near-maximal use of the available data for training \citep{AIROLA20111828}.

Our goal is to predict unseen rows and columns of the covariance at new redshift locations. For a fixed set of hyperparameters, LOO evaluates exactly this capability by holding out one redshift at a time and assessing the predicted row and column for that redshift. It also avoids arbitrary fold splits and keeps almost all information in the training set, which is important with only $m=15$ published redshifts.

Let $S\in\mathbb{R}^{m\times m}$ be the published $15\times15$ covariance on redshifts $z_{1:m}$. For each $i\in\{1,\dots,m\}$ we: (i) leave out $z_i$; (ii) train the model on the remaining $m-1$ redshifts; (iii) predict the full $m\times m$ covariance $\hat{\Sigma}^{(i)}$; and (iv) score the $i$th row and column against $S$ by root-mean-squared error (RMSE, counting the diagonal once). This produces one LOO score $s_i$ per redshift. We then aggregate the $m$ scores using the median and obtain the LOO median RMSE
\begin{equation}
    s_{\mathrm{LOO}}=\operatorname{median}_{i=1,\dots,m}\; s_i .
    \label{eq:selection_hyper1}
\end{equation}

Because results can vary slightly each time the code is executed with the same hyperparameter combination, we repeat the entire LOO procedure five times with different random seeds. Each run $r\in\{1,\dots,5\}$ produces a single score $s_{\mathrm{LOO},r}$. We then compute the mean
\begin{equation}
    \bar{s}=\frac{1}{5}\sum_{r=1}^{5} s_{\mathrm{LOO},r} .
    \label{eq:selection_hyper2}
\end{equation}
We select the hyperparameter combination that minimizes $\bar{s}$, which represents the best average performance with reduced sensitivity to lucky or unlucky seeds. The selected hyperparameters are reported in Table~\ref{tab:selected_hparams}. After choosing the winning hyperparameters, we identify the median run $r^\star$, the run whose $s_{\mathrm{LOO},r}$ is the middle value among $\{s_{\mathrm{LOO},1},\dots,s_{\mathrm{LOO},5}\}$. And use that run to generate the final simulated $33\times 33$ covariance matrix. The bottom-right panel of Figure~\ref{fig:top_publishedCM_bottom_PDCovNetCMs} shows the result. This rule, mean-to-select and median-to-report, reduces randomness while avoiding optimistic bias from a single best seed.

\begin{table}
\centering
\setlength{\tabcolsep}{8pt}
\renewcommand{\arraystretch}{1.2}
\caption{Selected hyperparameters.}
\label{tab:selected_hparams}
\begin{tabular}{lc}
\hline
Hyperparameter & Value \\
\hline
epochs & 3250 \\
feature dimension & 4 \\
hidden width (neurons per layer) & 128 \\
hidden depth (hidden layer) & 3 \\
learning rate & $10^{-3}$ \\
\hline
\end{tabular}
\end{table}

\subsection{GP baseline}
\label{subsec:GP baseline}
Section \ref{subsec:PD-CovNet} introduced PD-CovNet, which expands a published $15\times 15$ covariance matrix to a full $33\times 33$ covariance matrix. Its performance, however, has not yet been quantified. Given the low-data setting, nonparametric methods, especially Gaussian Processes, are appropriate as a reference. Here we construct a GP baseline that generates the full $33\times 33$ covariance and compare the matrices simulated by PD-CovNet and by the GP in Section~\ref{subsec:Five-fold cross-validation}.

We model the covariance matrix as a smooth bivariate function of two redshifts. Let $\mathbf Z_{15}=\{z_1,\ldots,z_{15}\}$ denote the redshifts at which the published $15\times15$ covariance $\boldsymbol{\Sigma}_{\rm pub}$ is available, and let $\mathbf Z_{33}=\{z'_1,\ldots,z'_{33}\}$ denote the target redshift list on which we seek the full matrix. A Gaussian Process \citep{2006gpml.book.....R} is fitted to the entries of $\boldsymbol{\Sigma}_{\rm pub}$ using the pairs $(z_i,z_j)\in\mathbf Z_{15}\times\mathbf Z_{15}$ and is then evaluated on all pairs $(z_i,z_j)\in\mathbf Z_{33}\times\mathbf Z_{33}$ to produce a symmetric, positive semidefinite $33\times33$ covariance matrix. Redshifts are standardized using statistics computed on $\mathbf Z_{15}$, and the same transformation is applied to $\mathbf Z_{33}$. For each pair $(z_i,z_j)$ we use the mean redshift and the separation
\begin{equation}
  \mu_{ij}=\frac{1}{2}(z_i+z_j),
  \label{eq:GPbaseline1}
\end{equation}
\begin{equation}
    \delta_{ij}=|z_i-z_j|,
    \label{eq:GPbaseline2}
\end{equation}
which follows the midpoint-lag (average redshift and separation) reparameterization used in locally stationary covariance modeling and in pair statistics in astronomy \citep{2011arXiv1109.4174D, Coil2013}. Both impose independent smoothness along each coordinate.

From the published $15\times15$ covariance matrix we form one training example for each distinct pair $(z_i,z_j)$: the feature is $(\mu_{ij},\delta_{ij})$ and the target is $\Sigma_{\mathrm{pub},ij}$. Collecting these gives
\begin{equation}
    \mathbf X=\{(\mu_{ij},\delta_{ij})\},
    \label{eq:GPbaseline3}
\end{equation}
\begin{equation}
    \mathbf y=\{\Sigma_{\mathrm{pub},ij}\}.
    \label{eq:GPbaseline4}
\end{equation}
A zero mean GP prior is placed on $s(\mu,\delta)$ with an anisotropic \text{Mat\'{e}rn} $(\nu = 5/2)$ kernel \citep{2023ApJS..266...27Z}:
\begin{equation}
k_\theta\!\big((\mu,\delta),(\mu',\delta')\big)
=\sigma_f^2\!\Big(1+\sqrt{5}\,r+\frac{5}{3}r^2\Big)\exp(-\sqrt{5}\,r),
\label{eq:GPbaseline5}
\end{equation}
\begin{equation}
    r^2=\frac{(\mu-\mu')^2}{l_\mu^2}+\frac{(\delta-\delta')^2}{l_\delta^2}.
    \label{eq:GPbaseline6}
\end{equation}
We assemble the training kernel matrix
\begin{equation}
\mathbf K_\theta=\big[k_\theta(\mathbf X_a,\mathbf X_b)\big]_{a,b},
\label{eq:GPbaseline7}
\end{equation}
where $\mathbf X_a$ and $\mathbf X_b$ denote the feature vectors for two distinct redshift pairs from $\mathbf Z_{15}$. We use a numerically stable version for inversion
\begin{equation}
    \tilde{\mathbf K}=\mathbf K_\theta+\sigma_n^2\mathbf I.
    \label{eq:GPbaseline8}
\end{equation}
The term $\sigma_n^2\mathbf I$ is a tiny diagonal addition used only for numerical stability of the code.
Hyperparameters $\theta=\{\ell_\mu,\ell_\delta,\sigma_f,\sigma_n\}$ are obtained by maximizing the log marginal likelihood
\begin{equation}
\log \mathcal L(\theta)
=-\frac12\,\mathbf y^{\!\top}\tilde{\mathbf K}^{-1}\mathbf y
-\frac12\log|\tilde{\mathbf K}|
-\frac{n}{2}\log(2\pi).
\label{eq:GPbaseline9}
\end{equation}

\begin{figure*}
    \centering
    \includegraphics[width=0.8\linewidth]{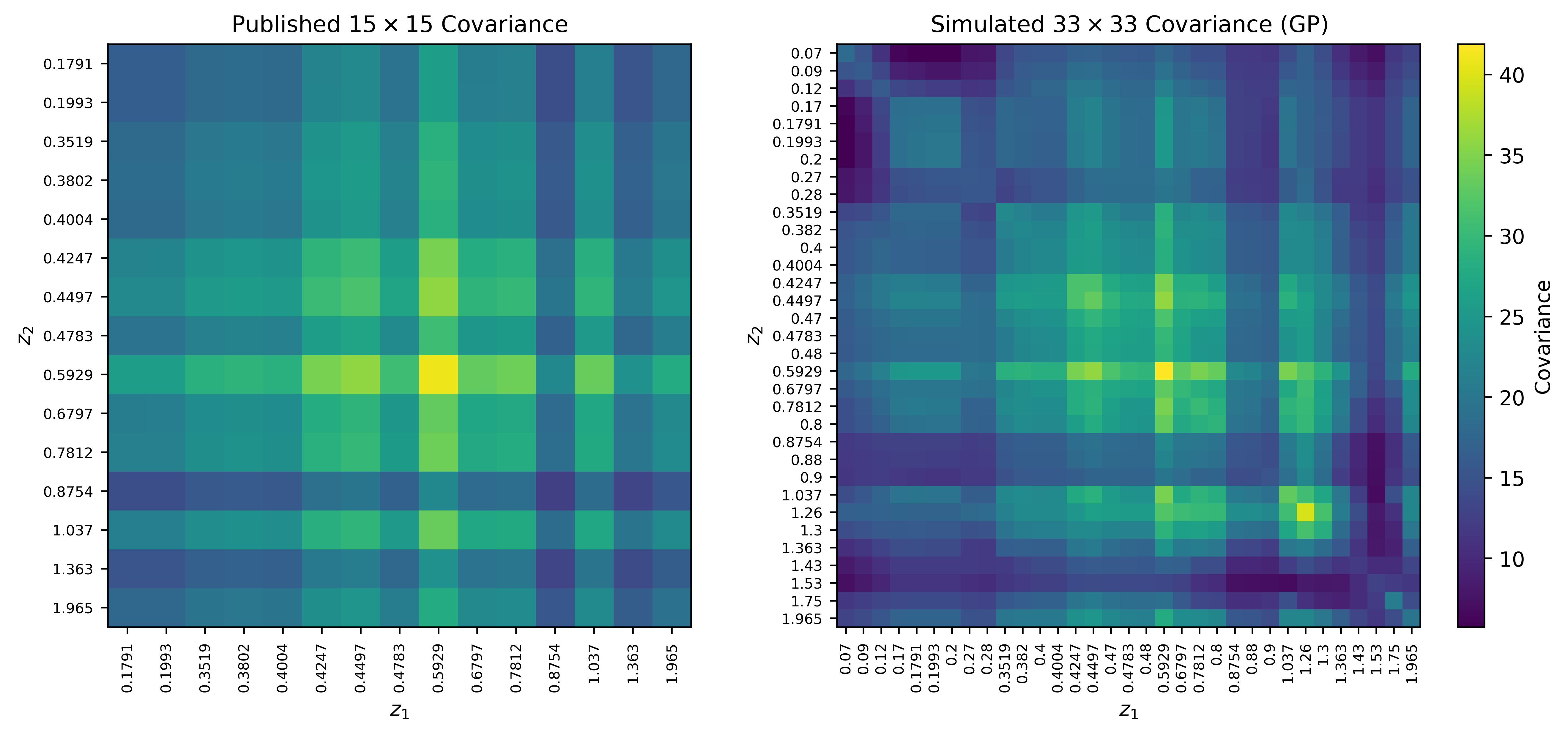}
    \caption{The published covariance matrix (left panel) and the GP-simulated covariance matrix (right panel). The left panel shows the published $15\times15$ covariance matrix $\boldsymbol{\Sigma}_{\rm pub}$ on the redshift set $\mathbf Z_{15}$. The right panel shows the simulated $33\times33$ covariance matrix $\boldsymbol{\Sigma}_{33}$ on $\mathbf Z_{33}$ obtained by fitting a Gaussian Process with an anisotropic \text{Mat\'{e}rn} $(\nu = 5/2)$ kernel to $\boldsymbol{\Sigma}_{\rm pub}$ and evaluating on $\mathbf Z_{33}$. Both panels use the same colormap with a single shared colorbar. The $x$- and $y$-axes are the redshifts $z_1$ and $z_2$, respectively.
}
    \label{fig:GP_simulatedCM_shared_cbar}
\end{figure*}

For each $(z_i,z_j)\in\mathbf Z_{33}\times\mathbf Z_{33}$, construct $(\mu_{ij},\delta_{ij})$ with the same standardization as in training and compute the predictive mean
\begin{equation}
    \tilde\Sigma_{ij} = \mathbf k\!\big((\mu_{ij},\delta_{ij}),\mathbf X\big)\,\tilde{\mathbf K}^{-1}\mathbf y.
    \label{eq:GPbaseline10}
\end{equation}
Stacking the entries $\tilde\Sigma_{ij}$ yields the matrix $\tilde{\boldsymbol\Sigma}_{33}$. We first enforce symmetry by averaging with the transpose
\begin{equation}
  M  = \tfrac{1}{2}\!\left(\tilde{\boldsymbol\Sigma}_{33}
  + \tilde{\boldsymbol\Sigma}_{33}^{\top}\right).
  \label{eq:GPbaseline11}
\end{equation}
To ensure positive semidefiniteness, we then project $M$ to the nearest PSD matrix using the Higham method \citep{10.1093/imanum/22.3.329}, yielding the final simulated covariance matrix shown in Figure~\ref{fig:GP_simulatedCM_shared_cbar}.

\subsection{Five-fold cross-validation}
\label{subsec:Five-fold cross-validation}
In Section~\ref{subsec:PD-CovNet}, we introduced PD-CovNet, which simulates a $33\times33$ covariance matrix from the published $15\times15$ covariance matrix. In this Section, we evaluate how accurately PD-CovNet reconstructs the covariance at unseen redshift pairs. To contextualize performance, we include a GP baseline in Section \ref{subsec:GP baseline}. Because we have only one $15\times15$ block, this is a low-data setting; a nonparametric GP often performs well in such low-data regimes. The GP is trained on the same block and evaluated on the same $33\times33$ covariance matrix. We then compare PD-CovNet and GP under identical five-fold cross-validation over redshifts, reporting fold-averaged accuracy and variability. This design yields a fair, cross-validation comparison on the $33\times33$ covariance matrix simulation task.

Cross-validation \citep{10.1111/j.2517-6161.1974.tb00994.x, 10.1111/j.2517-6161.1977.tb01603.x} is a model-assessment strategy for estimating how a fitted model will perform on unseen data. The dataset is repeatedly split into two disjoint parts: one part is used to train the model and the other is used only to evaluate it. By rotating these splits, we obtain an out-of-sample estimate of prediction error and its variability-especially valuable in our covariance-by-redshift setting with limited data. In $k$-fold cross-validation \citep{10.5555/1643031.1643047, 2018ApJ...869..167S}, the dataset is randomly partitioned into $k$ equal-sized subsamples ("folds"). At iteration $i=1,\dots,k$, fold $i$ is held out as the test set while the model is fitted on the remaining $k-1$ folds; the $k$ test scores are then averaged. Thus, each observation is used exactly once for testing and $k-1$ times for training, yielding a stable out-of-sample estimate. In our experiments we set $k=5$, producing five folds of three redshifts.

We first sort the 15 redshifts of the published $15\times15$ covariance matrix in ascending order and partition them into five equal-sized folds of three redshifts each. For a given fold $f$, the three redshifts in that fold are held out for testing, and the remaining twelve define the $12\times12$ training submatrix $S_{\text{train}}^{(f)}$. To ensure stable optimization and consistent scaling across folds, we standardize the redshift coordinate within each fold. The same transform is then applied to all 15 redshifts when making predictions. This preserves a strict train/test separation while improving numerical conditioning for both PD-CovNet and the GP.

For each fold, both models, PD-CovNet and the GP, are trained on $S_{\text{train}}^{(f)}$. PD\text{-}CovNet is fitted by minimizing the multivariate normal log-likelihood on the $12\times12$ block and then used to produce a full $15\times15$ prediction $\hat S^{(f)}$. By construction, $\hat S^{(f)}$ is symmetric and positive-definite. The GP baseline is fitted on the same training block via marginal-likelihood optimization with an anisotropic \text{Mat\'{e}rn} ($\nu=5/2$) kernel and then used to predict the full $15\times15$ prediction $\hat S^{(f)}$. The prediction using GP is subsequently symmetrized and PSD matrix to ensure a valid covariance.

From each full prediction, denoted $\hat S^{(f)}$, we extract the predicted held-out $3\times3$ block corresponding to the test redshifts and compare it to the true held-out block using root-mean-squared error (RMSE) and mean absolute error (MAE). Let the true held-out block and its prediction be $S_{\text{test}}^{(f)}$ and $\hat S_{\text{test}}^{(f)}$, and define the error matrix
\begin{equation}
    E^{(f)}=\hat S_{\text{test}}^{(f)}-S_{\text{test}}^{(f)}.
    \label{eq:fivefold-err}
\end{equation}
All metrics are computed on the $3\times3$ block, nine entries in total. RMSE and MAE are defined as
\begin{equation}
    \mathrm{RMSE}^{(f)}=\sqrt{\frac{1}{9}\sum_{i=1}^{3}\sum_{j=1}^{3}\big(E^{(f)}_{ij}\big)^2},
    \label{eq:fivefold-rmse}
\end{equation}
\begin{equation}
    \mathrm{MAE}^{(f)}=\frac{1}{9}\sum_{i=1}^{3}\sum_{j=1}^{3}\big|E^{(f)}_{ij}\big|.
    \label{eq:fivefold-mae}
\end{equation}
We repeat this procedure across all five folds. For both PD\text{-}CovNet and the GP, we report the mean and standard deviation of each metric, as shown in Table~\ref{tab:5fold-cv-results}. We use the same ordered folds, preprocessing, and scoring protocol for PD-CovNet and the GP. Choosing contiguous $z$-grouped folds (rather than random splits) respects the one-dimensional structure in redshift and yields a harder, more realistic generalization test.

\begin{table}  
\centering
\setlength{\tabcolsep}{8pt}
\renewcommand{\arraystretch}{1.2}
\caption{Five-fold cross-validation results (mean $\pm$ std over folds).}
\label{tab:5fold-cv-results}
\begin{tabular}{lcc}
\hline
Model & RMSE & MAE \\
\hline
PD-CovNet & $18.78 \pm 11.45$ & $17.78 \pm 10.54$ \\
GP        & $64.51 \pm 69.17$ & $59.52 \pm 61.90$ \\
\hline
\end{tabular}
\end{table}

From Table~\ref{tab:5fold-cv-results}, across five folds, PD-CovNet attains substantially lower errors than the GP baseline on both RMSE and MAE. Its errors are roughly 3 to 4 times lower than the GP's, and it exhibits much smaller fold-to-fold variability. The large GP standard deviations indicate instability under the contiguous $z$-grouped five-fold splits. Overall, PD-CovNet performs better than the GP for covariance simulation.

Additionally, as a brief sanity check, we simulate the $33\times33$ covariance matrix from the published $15\times15$ covariance matrix and then compare the simulated entries at the same redshift positions with the published $15\times15$ covariance matrix. The resulting RMSE and MAE are reported in Table~\ref{tab:Sanity check results}. PD-CovNet attains much smaller RMSE and MAE than the GP, which is consistent with the five-fold cross-validation results. Together, these findings indicate that PD-CovNet better simulates the $33\times33$ covariance matrix from the published $15\times15$ covariance matrix data than the GP.

\begin{table}  
\centering
\setlength{\tabcolsep}{8pt}
\renewcommand{\arraystretch}{1.2}
\caption{Sanity check results.}
\label{tab:Sanity check results}
\begin{tabular}{lcc}
\hline
Model & RMSE & MAE \\
\hline
PD-CovNet & 0.735 & 0.713 \\
GP        & 1.132 & 0.865 \\
\hline
\end{tabular}
\end{table}

\section{Constraining $\mathrm{H}_0$ Using the CC Dataset}
\label{sec:Constraining H0 Using the CC Dataset}
This section assesses how PD-CovNet hyperparameter choices and CC data configurations with different covariance matrices affect the constraint on $\mathrm{H}_0$. Using the CC dataset in Table~\ref{tab:cc_data}, we simulate a $33\times33$ covariance (PD33) with PD-CovNet trained on the published $15\times15$ covariance. And compare four configurations: NoCov (no covariance), Cov15 (published $15\times15$ covariance), PD33 (PD-CovNet $33\times33$), and PD33Alt denotes an alternative hyperparameter setting). To minimize the influence of the chosen constraint method on the constrained $\mathrm{H}_0$ results, this research employs two complementary approaches: EMCEE and GP.

\subsection{EMCEE}
\label{subsec:EMCEE}

In this Section, we utilize the \href{https://emcee.readthedocs.io/en/stable/index.html}{EMCEE} to constrain the Hubble constant, $\mathrm{H}_0$, based on the CC dataset while incorporating a covariance matrix to account for correlations between data points. This approach ensures robust and reliable parameter estimation. To model the relationship between the Hubble parameter, $H(z)$, and the cosmological parameters, we adopt the flat $\Lambda$CDM cosmological model. The corresponding Friedmann equation is expressed as:
\begin{equation}
    H(z) = \mathrm{H}_0 \sqrt{\Omega_M (1+z)^3 + (1-\Omega_M)},
    \label{eq:emcee1}
\end{equation}
where $H(z)$ represents the Hubble parameter at redshift $z$, $\mathrm{H}_0$ is the Hubble constant, and $\Omega_M$ is the matter density parameter. 

The CC dataset provides the observational values $H_{\mathrm{obs}}(z)$, their associated uncertainties $\sigma_{\mathrm{obs}}$, and the redshift values $z$. Additionally, the covariance matrix $\mathrm{Cov}$ accounts for the correlations between these observations. To constrain $\mathrm{H}_0$ and $\Omega_M$, we employ Bayes' theorem, which expresses the relationship between the posterior probability of the parameters, the likelihood of the observed data, and the prior distribution:  
\begin{equation}
    P(\mathrm{H_0,\Omega_M}|H_{\mathrm{obs}}) = \frac{P(H_{\mathrm{obs}}|\mathrm{H_0,\Omega_M}) P(\mathrm{H_0,\Omega_M})}{P(H_{\mathrm{obs}})},
    \label{eq:emcee2}
\end{equation}
where $P(H_{\mathrm{obs}} | \mathrm{H}_0, \Omega_M)$ is the likelihood function, which quantifies the probability of obtaining the observed data given specific values of $\mathrm{H}_0$ and $\Omega_M$. The term $P(\mathrm{H}_0, \Omega_M)$ represents the prior distribution, encapsulating prior knowledge or assumptions about these parameters before considering the data. The denominator, $P(H_{\mathrm{obs}})$, known as the evidence or marginal likelihood, serves as a normalization factor to ensure that the posterior probability $P(\mathrm{H}_0, \Omega_M | H_{\mathrm{obs}})$ is properly scaled.

The likelihood function is expressed as \citep{2011ApJ...730...74M,2021ApJS..254...43W}:
\begin{equation}
    \mathcal{L} = \exp{\left(-\frac{1}{2} \chi^2 \right)},
    \label{eq:emcee3}
\end{equation}
where the $\chi^2$ statistic quantifies the goodness of fit between the model and the observations. The $\chi^2$ is defined as:
\begin{equation}
    \chi^2 = \bm{r}^T \, \mathrm{Cov}^{-1} \, \bm{r},
    \label{eq:emcee4}
\end{equation}
with the residual vector $\bm{r}$ given by:
\begin{equation}
    r_i = H_{\mathrm{obs},i} - H_{\mathrm{model},i}.
    \label{eq:emcee5}
\end{equation}
Here, $H_{\mathrm{obs},i}$ and $H_{\mathrm{model},i}$ are the observed and model-predicted Hubble parameter values, respectively, and $\mathrm{Cov}^{-1}$ is the inverse of the covariance matrix. This formulation ensures that the $\chi^2$ accounts for both the individual variances and the correlations between the observations.

Using EMCEE, we sample the posterior distribution of $\mathrm{H}_0$ and $\Omega_M$ to derive their constraints, effectively incorporating the effects of the covariance matrix into the analysis. This method provides robust and statistically rigorous estimates of the parameters. To evaluate the sensitivity of the constrained $\mathrm{H}_0$ to PD-CovNet hyperparameters used to simulate the covariance, we generate two covariance matrices with different hyperparameters: the selected configuration, PD33 (Figure~\ref{fig:top_publishedCM_bottom_PDCovNetCMs}, bottom right), and an alternative, PD33Alt (Figure~\ref{fig:top_publishedCM_bottom_PDCovNetCMs}, bottom left). We then constrain $\mathrm{H}_0$ with EMCEE under four CC data configurations: NoCov (no covariance), Cov15 (published $15\times15$), PD33, and PD33Alt, to assess how the covariance itself influences the constraint. The resulting estimates are reported in Table~\ref{tab: Constraints on H0}. We compare and discuss these results in Section~\ref{subsec: Constraints on H0}.

\subsection{GP}
\label{subsec:GP}

In this section, to minimize the impact of the constraint method, we hold the CC data and covariance fixed and replace EMCEE with a GP to isolate method effects on the $\mathrm{H}_0$ constraint. GP regression provides a model-independent method for reconstructing the CC dataset and estimating $\mathrm{H}_0$. Following \cite{2006gpml.book.....R}, GP is a non-parametric approach that models data as a multivariate Gaussian distribution, enabling the interpolation of unknown values without assuming an explicit functional form. This makes it particularly useful for reconstructing the Hubble parameter $H(z)$ as a function of redshift $z$, where $\mathrm{H}_0$ is determined from the reconstructed value of $H(z)$ at $z = 0$. In this study, we enhance the standard GP approach by incorporating the full covariance matrix of the CC dataset, which is simulated using a PD-CovNet. Unlike previous studies \citep{2021ApJ...915..123S, 2023ApJS..266...27Z} that assumed independent data points with only diagonal uncertainties, our approach effectively captures correlations between redshift measurements. This provides a more comprehensive statistical treatment of the data and improves the robustness of the resulting $\mathrm{H}_0$ constraint.  

To model the observed CC data, given as ($z_i, H_i, \sigma_i$), we assume that the errors $\sigma_i$ follow a Gaussian distribution. The data vector $\bm{H}$ is then described by
\begin{equation}
    \bm{H} \sim \mathcal{N} ( \bm{\mu}, K(\bm{Z},\bm{Z}) + C), 
    \label{eq:gp1}
\end{equation}
where $\mathcal{N}$ represents a Gaussian Process, $\bm{\mu}$ is the mean function, and $K(\bm{Z},\bm{Z})$ is the covariance function that captures the relationships between data points. Here, $C$ is the full covariance matrix of the CC dataset, which accounts for both statistical and systematic uncertainties. It is defined in Equation~(\ref{eq:data2}). The covariance between two data points, $H(z_i)$ and $H(z_j)$, is given by a covariance function, also known as a kernel. There are multiple choices for the kernel function \citep{2023ApJS..266...27Z}, but in this study, we adopt the widely used Squared Exponential covariance function \citep{2012JCAP...06..036S,2021ApJS..254...43W,2021ApJ...915..123S}
\begin{equation}
    k( z_i, z_j ) = \sigma^2_{f} \exp{\left( -\frac{( z_i - z_j )^2}{2l^2} \right)},
    \label{eq:gp2}
\end{equation}
where $l$ is the characteristic length scale controlling the smoothness of the function, and $\sigma_f$ is the signal variance determining the amplitude of the fluctuations. These hyperparameters are optimized during the GP fitting process.

To extend the GP model to new redshift points $\bm{Z^{\ast}}$, we define a corresponding Gaussian vector
\begin{equation}
    \bm{f^{\ast}} \sim \mathcal{N} ( \bm{\mu^{\ast}}, K(\bm{Z^{\ast}},\bm{Z^{\ast}})).
    \label{eq:gp3}
\end{equation}
Combining Equations (\ref{eq:gp1}) and (\ref{eq:gp3}) in a joint Gaussian distribution \citep{2012JCAP...06..036S}, we obtain
\begin{equation}
    \begin{bmatrix}
        \bm{H}  \\
        \bm{f^{\ast}}  \\
    \end{bmatrix}
    \sim \mathcal{N}  
    \left( 
    \begin{bmatrix}
        \bm{\mu} \\
        \bm{\mu^{\ast}} \\
    \end{bmatrix}
    ,
    \begin{bmatrix}
        K(\bm{Z},\bm{Z}) + C & K(\bm{Z},\bm{Z^{\ast}}) \\
        K(\bm{Z^{\ast}},\bm{Z}) & K(\bm{Z^{\ast}},\bm{Z^{\ast}}) \\
    \end{bmatrix}    
    \right).
    \label{eq:gp4}
\end{equation}
From this, we derive the mean and covariance of the reconstructed function \citep{2006gpml.book.....R,2012JCAP...06..036S}
\begin{equation}
    \bm{\bar{f^{\ast}}} = \bm{\mu^{\ast}} + K(\bm{Z^{\ast}},\bm{Z}) [K(\bm{Z},\bm{Z}) + C]^{-1} (\bm{H} - \bm{\mu}),
    \label{eq:gp5}
\end{equation}
\begin{align}
    \operatorname{cov}(\bm{f^{\ast}}) &= K(\bm{Z^{\ast}}, \bm{Z})
    - K(\bm{Z^{\ast}}, \bm{Z}) [K(\bm{Z}, \bm{Z}) + C]^{-1} \notag \\
    &\quad - K(\bm{Z}, \bm{Z^{\ast}}).
    \label{eq:gp6}
\end{align}
The hyperparameters $l$ and $\sigma_f$ are determined by maximizing the log marginal likelihood
\begin{align}
    \ln \mathcal{L} &= - \frac{1}{2} ( \bm{H} - \bm{\mu} )^{T} [K(\bm{Z}, \bm{Z}) + C]^{-1} ( \bm{H} - \bm{\mu} ) \notag \\
    &\quad - \frac{1}{2} \ln |K(\bm{Z}, \bm{Z}) + C| - \frac{n}{2} \ln 2\pi.
    \label{eq:gp7}
\end{align}

Following this procedure, we reconstruct $H(z)$ and obtain $\mathrm{H}_0$ by evaluating the reconstructed function at $z=0$. For implementation, we use the Gaussian process regression package GAPP (Gaussian Processes in Python) developed by \cite{2012JCAP...06..036S}. The inclusion of the full covariance matrix provides an improved statistical treatment and leads to a more robust constraint on $\mathrm{H}_0$ from the CC dataset. 

In this section, to minimize the impact of the constraint method, we repeat the EMCEE analysis step-for-step but replace EMCEE with GP. We hold the CC dataset and each covariance configuration fixed: NoCov, Cov15, PD33, and PD33Alt (with PD33/PD33Alt generated by PD-CovNet as described in Section \ref{subsec:PD-CovNet}). And re-estimate $\mathrm{H}_0$ using GP. Results are reported alongside the EMCEE outcomes in Table~\ref{tab: Constraints on H0}, enabling a direct, method-only comparison; the implications are discussed in Section \ref{subsec: Constraints on H0}.

\subsection{Constraints on $\mathrm{H}_0$}
\label{subsec: Constraints on H0}
This section assesses how the CC dataset specification, covariance modeling, and constraint method affect the estimate of $\mathrm{H}_0$. First, to quantify the impact of the CC covariance matrix on $\mathrm{H}_0$, we analyze four CC data configurations:
(i) CC without a covariance matrix;
(ii) CC with the published $15\times15$ covariance;
(iii) CC with a simulated $33\times33$ covariance generated by PD-CovNet using the hyperparameters selected in Section~\ref{subsubsec:selection of hyperparameters}; and
(iv) the same as (iii) but with an alternative hyperparameter combination to assess the sensitivity of the $\mathrm{H}_0$ constraint to PD-CovNet hyperparameters (see Figure~\ref{fig:top_publishedCM_bottom_PDCovNetCMs}, bottom-left panel). We use these short labels (CC-NoCov, CC-Cov15, CC-PD33, CC-PD33Alt), corresponding to (i)-(iv), respectively, throughout tables. Second, to minimize method-specific artifacts in the inference itself, we estimate $\mathrm{H}_0$ with two methods, EMCEE and the GP, described in Sections~\ref{subsec:EMCEE} and \ref{subsec:GP}. All $\mathrm{H}_0$ constraints are summarized in Table~\ref{tab: Constraints on H0}.

\begin{table} 
\centering
\setlength{\tabcolsep}{8pt}
\renewcommand{\arraystretch}{1.2}
\caption{$\mathrm{H}_0$ from CC configurations via EMCEE and GP.}
\label{tab: Constraints on H0}
\begin{tabular}{lcc}
\hline
CC data configurations & $\mathrm{H}_0$\textsuperscript{a} via EMCEE & $\mathrm{H}_0$ via GP \\
\hline
CC-NoCov   & $67.92 \pm 3.10$ & $67.76 \pm 4.78$ \\
CC-Cov15   & $69.17 \pm 4.16$ & $67.22 \pm 4.72$ \\
CC-PD33    & $67.71 \pm 5.28$ & $67.89 \pm 5.97$ \\
CC-PD33Alt & $68.35 \pm 5.03$ & $68.06 \pm 5.69$ \\
\hline
\end{tabular}

\parbox{\linewidth}{\small
\textsuperscript{a} Values are posterior mean $\pm 1\sigma$; units km s$^{-1}$ Mpc$^{-1}$.}
\end{table}

To quantify how CC data configurations and constraint methods influence the constrained $\mathrm{H}_0$, we compare the $\mathrm{H}_0$ values in Table~\ref{tab: Constraints on H0} using two complementary measures: (i) statistical significance via the standardized difference $S$ (Gaussian $n$-$\sigma$ tension) with its associated two-sided $p$-value, and (ii) the precision ratio $\rho$ of the corresponding $1\sigma$ half-widths.

To compare two values $X=\mu_X\pm\sigma_X$ and $Y=\mu_Y\pm\sigma_Y$, we place the uncertainty on the comparison itself by defining the difference $\Delta = Y - X$ with mean $\mu_\Delta = \mu_Y - \mu_X$ and variance $\sigma_\Delta^2 = \sigma_X^2 + \sigma_Y^2$. We report the standardized difference $S$ between two independent values:
\begin{equation}
S \equiv \frac{|\mu_\Delta|}{\sigma_\Delta}.
\label{eq:Constraints on H0_1}
\end{equation}
Smaller $S$ (reported in $\sigma$ units) indicates closer agreement. Values well below $1\sigma$ imply no meaningful separation, while values of several~$\sigma$ indicate strong discrepancy. The two-sided $p$-value is the probability, assuming the two estimates have the same true value, that a discrepancy at least as large as the one observed would arise due to random variability. Under the usual independence approximation for the difference, the standardized difference $S$ is standard normal when the true difference is zero. The associated two-sided $p$-value is
\begin{equation}
p \;=\; 2\bigl[1-\Phi(S)\bigr],
\label{eq:Constraints on H0_2}
\end{equation}
where $\Phi$ denotes the cumulative distribution function of the standard normal distribution. This mapping simply expresses the $\sigma$-level effect size $S$ on a conventional significance scale. The $p$-value, $p$, is computed to judge how consistent two estimates are with being the same. Large $p$ ($\ge 0.10$) indicates little evidence of a real difference; $p \in (0.05,\,0.10)$ indicates weak evidence; and $p \le 0.05$ is typically treated as statistically significant evidence of a difference. Changes in precision are summarized by the 1$\sigma$ ratio
\begin{equation}
\rho = \frac{\sigma_Y}{\sigma_X}.
\label{eq:Constraints on H0_3}
\end{equation}
Thus $\rho>1$ indicates that $Y$ is less precise than $X$ by $(\rho-1)\times100\%$, and $\rho<1$ indicates greater precision.

Using the eight constrained $\mathrm{H}_0$ values in Table~\ref{tab: Constraints on H0}, we form all $\binom{8}{2}=28$ pairwise comparisons and evaluate both the standardized difference $S$ and the associated two-sided $p$-value. For $S$, the mean, median, standard deviation, minimum, and maximum are  $\overline{S}=0.095$, $ \mathrm{median}(S)=0.076,$, $\mathrm{sd}(S)=0.080$, $S_{\min}=0.004$, $ S_{\max}=0.310$, respectively. 18/28 pairs satisfy $S\le 0.10\sigma$, 27/28 satisfy $S\le 0.25\sigma$, and 28/28 satisfy $S\le 0.31\sigma$. The corresponding $p$-values are tightly concentrated near unity: $\overline{p}=0.924$, $\mathrm{median}(p)=0.939$, $\mathrm{sd}(p)=0.063$, $p_{\min}=0.757$, and $p_{\max}=0.996$; 28/28 pairs have $p\ge 0.75$. The largest observed shift is only $S=0.31\sigma$ (corresponding to $p=0.757$). These results are fully consistent: small $S$ values correspond to large $p$ values, indicating that across CC data configurations and between EMCEE and GP there is no statistically meaningful shift in the central value of the constrained \(\mathrm{H}_0\).

From Table~\ref{tab:rho-summary}, for a fixed CC dataset, changing the constraint method from EMCEE to GP has only a modest effect once covariance is modeled: uncertainties increase by about 13\% for Cov15, PD33, and PD33Alt. The only clear exception is the NoCov case, where GP's uncertainty is 54.2\% larger than EMCEE's. Thus, the constraint method does impact precision: EMCEE is uniformly more precise than GP. By contrast, within a given method the choice of CC data configuration clearly affects precision. Under EMCEE, precision degrades from NoCov to Cov15 to PD33, with imprecision increasing by 34\% from NoCov to Cov15 and by 27\% from Cov15 to PD33. Under GP, Cov15 and NoCov are essentially indistinguishable in precision ($\approx$1\% difference), while PD33 is 27\% less precise than Cov15. Overall, both methods agree that introducing more covariance reduces precision; the only exception is that NoCov and Cov15 are nearly identical under GP.

\begin{table*}
\centering
\caption{Precision ratios $\rho=\sigma_Y/\sigma_X$ for selected pairs within each method.}
\label{tab:rho-summary}
\begin{tabular}{lcccc}
\hline
Method & Cov15/NoCov  & PD33/Cov15  & PD33/NoCov  & PD33Alt/PD33 \\
\hline
EMCEE & $1.342$ (+34.2\% \textsuperscript{a}) & $1.269$ (+26.9\%) & $1.703$ (+70.3\%) & $0.953$ ($-4.7$\%) \\
GP    & $0.987$ ($-1.3$\%) & $1.265$ (+26.5\%) & $1.249$ (+24.9\%) & $0.953$ ($-4.7$\%) \\
\hline
\multicolumn{5}{l}{\footnotesize \textsuperscript{a} Percent shows change in imprecision: $(\rho-1)\times100\%$.}
\end{tabular}
\end{table*}

Finally, hyperparameters affect the learned CC covariance by trading smoothness against detail and by controlling numerical stability. In our tuning, epochs with early stopping influence the smoothness of the simulated covariance by controlling how much detail the model captures during training. Fewer epochs lead to smoother covariances, while training longer adds finer detail but may introduce noise. Feature dimension $r$ sets the effective rank of the covariance, influencing how much structure is preserved or smoothed. Hidden width and depth define the network's capacity to capture complex correlations in the data. Learning rate governs the stability of training, determining how quickly the network converges and how stable the covariance matrix estimate is. 

These hyperparameters influence how the network learns to represent and capture covariance, impacting its final structure, which in turn affects the precision of the constrained $\mathrm{H}_0$. The $S$ and $p$-value results for all $\mathrm{H}_0$ entries in Table~\ref{tab: Constraints on H0} show uniformly small $S$ and large $p$, indicating no statistically meaningful shift in the central value of the constrained \(\mathrm{H}_0\). But the uncertainty of $\mathrm{H}_0$ depends on the covariance configurations and PD-CovNet's hyperparameters. Because Table~\ref{tab: Constraints on H0} demonstrates that different covariance matrices lead to different $\mathrm{H}_0$ uncertainties, altering hyperparameters, which alters the covariance, will also alter the precision. Empirically, PD33Alt is about 4.7\% more precise than PD33 under both EMCEE and GP. Thus, although the sensitivity is weak, it is non-negligible, and careful hyperparameter selection is warranted when simulating the CC covariance with PD-CovNet. A systematic hyperparameter study is beyond the scope of this work and is left for future research.

\section{Conclusion}
\label{sec:Conclusion}

In this study, we investigated the impact of incorporating a simulated covariance matrix on constraining the Hubble constant, $\mathrm{H}_0$, using the CC dataset. Recognizing that previous studies either neglected off-diagonal elements \citep{2023PDU....3901147N, 2023ApJS..266...27Z} or used an incomplete covariance matrix \citep{2024ApJS..270...23Z, 2024ApJ...972...14N}, we addressed this gap by employing PD-CovNet to simulate the full $33 \times 33$ covariance matrix. Since the full covariance matrix of the CC dataset has not been explicitly published, we based our simulation on the $15 \times 15$ covariance matrix from \citet{2020ApJ...898...82M}. This simulation is grounded in the working hypothesis that the covariance matrix follows a generalizable pattern that PD-CovNet can learn, which we empirically validate via redshift-grouped five-fold cross-validation and by adding a GP baseline for comparison (see Section \ref{sec:Simulate Covariance Matrix}).

During the simulation of the covariance matrix, we observed that the hyperparameters materially affect the simulated covariance matrix, which were selected via LOO cross-validation in Section \ref{subsubsec:selection of hyperparameters}. In Section \ref{subsec:GP baseline}, a nonparametric GP baseline was trained on the same published $15 \times 15$ covariance matrix for reference. Under identical five-fold cross-validation (Section~\ref{subsec:Five-fold cross-validation}) over redshift groups, PD-CovNet achieved substantially lower errors than the GP baseline on both RMSE and MAE, typically by a factor of roughly 3 to 4 (Table~\ref{tab:5fold-cv-results}). A sanity check that compares the simulated entries at the overlapping \(15\times15\) positions with the published matrix (Table~\ref{tab:Sanity check results}) confirms the same pattern: PD-CovNet attains much smaller RMSE and MAE than the GP baseline. Taken together, these results indicate that PD-CovNet is the more reliable generator of the \(33\times33\) covariance.

Section~\ref{sec:Constraining H0 Using the CC Dataset} evaluates how covariance choice and constraint method affect the constrained $\mathrm{H}_0$. Using four CC configurations (NoCov, Cov15, PD33, and PD33Alt) and two independent constraint methods (EMCEE and GP), we find that the constrained $\mathrm{H}_0$ values are mutually consistent: standardized differences are small and two-sided p-values are large across all pairwise comparisons, indicating no statistically meaningful shift in the central value of the constrained \(\mathrm{H}_0\). Precision, however, does depend on both covariance specification and method. EMCEE is uniformly more precise than GP once covariance is modeled (uncertainties increase by \(\approx13\%\) when moving from EMCEE to GP for Cov15, PD33, and PD33Alt; the only exception is NoCov, where GP is \(54.2\%\) less precise). Within a given method, the choice of covariance matters: under EMCEE, precision degrades from NoCov \(\rightarrow\) Cov15 \(\rightarrow\) PD33 (by \(34\%\) and then \(27\%\)), while under GP, Cov15 and NoCov are nearly identical (\(\approx1\%\) difference) but PD33 is \(27\%\) less precise than Cov15. Both constraint methods consistently show that adding covariance information decreases precision. Varying PD-CovNet hyperparameters has little effect on the mean $\mathrm{H}_0$ but modestly impacts precision (e.g. PD33Alt is \(\sim4.7\%\) more precise than PD33). Taken together, the findings underscore that careful covariance modeling is essential for reliable CC-based \(\mathrm{H}_0\) constraints.

Future work can improve upon this study in several ways. (1) Exploring more advanced neural network architectures could enhance the accuracy of covariance matrix simulations. (2) We will investigate the mechanism behind the EMCEE-GP precision gap, specifically, why GP yields systematically wider credible intervals despite similar central \(\mathrm{H}_0\) values. (3) Directly measuring $H(z)$ with an observationally derived covariance matrix rather than relying on a simulated one would allow for a direct comparison to validate the PD-CovNet-based approach. (4) Further studies should refine the assumptions behind the covariance matrix simulation, such as testing whether the PD-CovNet is truly learning a valid underlying pattern and exploring alternative statistical methods for inferring missing correlations. These improvements will help refine $\mathrm{H}_0$ constraints and enhance our understanding of cosmic expansion.

\section*{Acknowledgements}

We thank Yu-Chen Wang and Kang Jiao for their useful discussions. This work was supported by National SKA Program of China，No.2022SKA0110202 and the China Manned Space Program with grant No. CMS-CSST-2025-A01.

\section*{DATA AVAILABILITY}
All data included in this study are available upon request by contact with the corresponding author. 




\bibliographystyle{mnras}
\bibliography{sample631} 




\appendix

\section{Correction of the Hubble Parameter at $z=0.09$}
\label{appendix: The correction of H(0.09)}

The Hubble parameter at redshift $z = 0.09$ is often reported as $H(0.09) = 69 \pm 12 \, \mathrm{km \, s^{-1} \, Mpc^{-1}}$ in numerous studies \citep{2023arXiv230709501M, 2024ApJ...972...14N}. The original source of this data point is \citet{2003ApJ...593..622J}, where $69 \pm 12 \, \mathrm{km \, s^{-1} \, Mpc^{-1}}$ is used as an estimate of the Hubble constant $\mathrm{H}_0$. To accurately obtain the Hubble parameter at redshift $z = 0.09$, we must adjust this value using the cosmological model parameters provided in that study, as follows
\begin{equation}
H(z) = \mathrm{H}_0 \sqrt{\Omega_M (1+z)^3 + \Omega_{\Lambda}},
\label{eq:correction1}
\end{equation}
where $\mathrm{H}_0 = 69 \pm 12 \, \mathrm{km \, s^{-1} \, Mpc^{-1}}$ is the Hubble constant, $\Omega_M = 0.27$ is the matter density parameter, and $\Omega_{\Lambda} = 0.7$ is the dark energy density parameter.

To determine the $1 \sigma$ uncertainty in $H(0.09)$, we apply uncertainty propagation. For a general function $f = f(x_1, x_2, \ldots, x_n)$, the variance is given by
\begin{equation}
\sigma_f^2 = \sum_{i=1}^{n} \left( \frac{\partial f}{\partial x_i} \right)^2 \sigma_{i}^2 + \sum_{i=1}^{n} \sum_{j \neq i}^{n} \frac{\partial f}{\partial x_i} \frac{\partial f}{\partial x_j} \sigma_{ij},
\label{eq:correction2}
\end{equation}
where $\sigma_i$ is the standard deviation of $x_i$, and $\sigma_{ij}$ represents the covariance between the variables $x_i$ and $x_j$. The covariance $\sigma_{ij}$ is calculated as
\begin{equation}
\sigma_{ij} = \frac{1}{N} \sum_{k=1}^{N} (x_{ik} - \bar{x_i})(y_{jk} - \bar{y_j}),
\label{eq:correction3}
\end{equation}
where $x_{ik}$ is the $k$-th value of $x_i$, and $N$ is the total number $k$. Based on Equation~(\ref{eq:correction2}), the propagated uncertainty in $H(z)$ is given by
\begin{equation}
\sigma_{H(z)} = \sqrt{\Omega_M (1 + z)^3 + \Omega_\Lambda} \cdot \sigma_{\mathrm{H}_0}.
\label{eq:correction4}
\end{equation}
According to Equations (\ref{eq:correction1}) and (\ref{eq:correction4}), we obtain the updated value for the Hubble parameter at $z = 0.09$ as $H(0.09) = 70.7 \pm 12.3 \, \mathrm{km \, s^{-1} \, Mpc^{-1}}$.


\label{lastpage}
\end{CJK*}
\end{document}